\documentclass[smallextended]{svjour2}

\usepackage{mathtools}

\usepackage{mathptmx}

\usepackage{comment}

\begin{document}

\title{Path integral representation for stochastic jump processes with boundaries}

\author{Takashi Arai}

\institute{T. Arai \at
              Graduate School of informatics, Kyoto University, Kyoto 606-8501, Japan \\
              \email{arai-t@sys.i.kyoto-u.ac.jp}
}

\date{Received: date / Accepted: date}

\maketitle

\begin{abstract}
We propose a formalism to analyze discrete stochastic processes with finite-state-level $N$.
By using an $(N+1)$-dimensional representation of su(2) Lie algebra, we re-express the master equation to a time-evolution equation for the state vector corresponding to the probability generating function.
We found that the generating function of the system can be expressed as a propagator in the spin coherent state representation.
The generating function has a path integral representation in terms of the spin coherent state.
We apply our formalism to a linear Susceptible-Infected-Susceptible (SIS) epidemic model with time-dependent transition probabilities.
The probability generating function of the system is calculated concisely using an algebraic property of the system or a path integral representation.
Our results indicate that the method of analysis developed in the field of quantum mechanics is applicable to discrete stochastic processes with finite-state-level.
\end{abstract}

\maketitle

\section{Introduction}
Continuous time Markov processes with discrete variable have been used to model phenomena over a wide range of fields, such as biology, ecology and stoichiometry~\cite{Gardiner}.
Among them, the system with infinite-state-level represented by a birth-death process have been well studied, however one often encounters the situation where the state-level is finite, such as spike activity of neurons or chemical reaction systems.
Recently, a phase-transition like phenomenon caused by the stochastic noise as represented by the autocatalytic reaction system and the foraging behavior of ants has attracting attentions~\cite{McKane,Kaneko,Vallade}, where the qualitative behavior of the system changes drastically as the system size becomes small.
These stochastic processes are described by the master equation, however it is in general difficult to analyze the master equation directly.
Therefore, one has to resort to an approximation scheme to analyze the system.

A well-known approximation scheme to the master equation is the Fokker-Planck approximation~\cite{Kampen}.
This approximation is able to capture the noise-induced bimodality in the model of foraging behavior of ants~\cite{McKane}.
However, the Fokker-Planck approximation is a continuous approximation to intrinsically discrete variables, and it was pointed out that the method may fail to capture some kind of feature of the stochastic system~\cite{Grima}.
Another method, which retains discreteness of the system, is to use the creation and annihilation operators for boson, initiated by Doi and Peliti~\cite{Doi,Doi2,Peliti}.
In this formalism, various statistics can be represented by the boson coherent state path integral, and this formalism gives a framework to apply various methods developed in quantum mechanics.
In fact, the Doi-Peliti formalism was applied to a biochemical reaction system and models of spike activity of neurons~\cite{Cowan,Bressloff,Chow}.
However, the Doi-Peliti formalism is originally the method to treat the infinite level systems such as growth of bacteria or variations of animal populations, and it seems to be not appropriate to treat a finite-level system using this formalism.
In fact, thermodynamic limit was taken in the application to the spike activity of neurons, on the assumption that the number of neuron in each population is infinite~\cite{Cowan,Bressloff,Chow}.
Such a thermodynamic limit may again lose some features of the stochastic system, and the finite-size effect could not be estimated correctly.
Therefore, an approximation scheme which captures the finite-size effect appropriately is desired.

Therefore in this paper, we propose a method to treat discrete stochastic processes with finite-state-level.
For this purpose, we use a representation of su(2) Lie algebra.
We found that by employing an appropriate normalization of state vectors in Hilbert space the master equation of the system can be rewritten as a time evolution equation for the state vector corresponding to the probability generating function. 
Then, the generating function is expressed as a propagator of the spin coherent state.
We derive the spin coherent state path integral representation for the generating function.
A similar formulation was shown previously in Ref.~\cite{Tailleur}, however we use a different representation of su(2) algebra.
We study the linear Susceptible-Infected-Susceptible (SIS) epidemic model with arbitrary time-dependent transition probabilities as a simplest application to demonstrate the effectiveness of our method.
We employ two approach, Wei-Norman theorem and path integral representation, to calculate the generating function.

This paper is organized as follows.
In Sec. \ref{sec:algebra}, we show how the discrete stochastic system with finite-state-level can be re-expressed by using a representation of su(2) algebra. 
In Sec. \ref{sec:application}, we study the stochastic linear SIS epidemic model as a simplest application of our method.
Sec. \ref{sec:conclusion} is devoted to conclusion.

\section{Algebra of finite-level systems \label{sec:algebra} }
The system of our interest consists of single population with $N$ individuals.
Each individual can take two discrete states.
For example, the two states correspond to firing and quiescent states in the model of spike activity of neurons, and susceptible and infected states in epidemic model.
Thus, $N$ individuals are divided into two classes.
The state of the system at a given time $t$ is represented by a discrete variable $n(t)$ with $(N+1)$ state-level, which takes $n(t) \in \{0, 1, \dots, N \}$.
Time-evolution of the system is assumed to be a continuous time Markov process with asynchronous updating, that is, only one individual changes the state during infinitesimal time interval.
Given a state transition probability from $n$ to $n'$ per unit time $w(n \rightarrow n')$, the probability with the state $n$ at time $t$ is described by the following master equation,
\begin{equation}
\frac{\partial}{\partial t} P(n,t)= \sum_{n'} w(n'\rightarrow n) P(n',t) -w(n \rightarrow n') P(n,t).
\end{equation}
Expressing the excitation and de-excitation probabilities for each individual as $f(n)$ and $g(n)$ respectively, the master equation is expressed as
\begin{align}
\frac{\partial}{\partial t} P(n,t)=& \bigl[N-(n-1) \bigr] f(n-1) P(n-1,t)-(N-n)f(n) P(n,t) \notag \\
&+(n+1)g(n+1) P(n+1,t) -n g(n) P(n,t),
\end{align}
where we have defined $P(-1,t)=P(N+1,t)=0$.

It is in general difficult to analyze the master equation directly.
Therefore, we first consider to rewrite the master equation to more tractable form.
In fact, by using a representation of su(2) algebra, the master equation can be rewritten as a time evolution equation for the state vector.
Though su(2) algebra is very familiar in the field of quantum mechanics, we once again illustrate the algebra here for the purpose of emphasizing the difference to the previous study~\cite{Tailleur} and quantum mechanical systems.

su(2) algebra is defined by the three generators satisfying the following commutation relations,
\begin{equation}
\begin{split}
& [J_+, J_-] = 2 J_z, \\
& [J_z, J_{\pm}] = \pm J_{\pm},
\end{split}
\end{equation}
where $J_+$ and $J_-$ are called raising and lowering operators.
Next, we introduce the lowest-weight state $|0 \rangle \equiv |J, -J\rangle$, which has a normalization $\langle 0| 0\rangle=1$ and satisfies the action $J_- |0\rangle =0$.
The parameter $J=N/2$ takes an integer or a half-integer.
Here, we define the following $(N+1)$-dimensional orthogonal vectors $|n\rangle$,
\begin{equation}
|n\rangle =\frac{(N-n)!}{N!} (J_+)^n |0\rangle.
\end{equation}
It should be noted that the inner-product of this state is not normalized to one, unlike that in quantum mechanics.
This normalization is crucial that one applies the su(2) algebra to the stochastic systems.
Then, the action of the operators $J_{\pm}$ and $J_z$ to the state $|n\rangle$ is respectively given by
\begin{equation}
\begin{split}
&J_+ |n\rangle= (N-n) |n+1\rangle, \\
&J_- |n\rangle= n |n-1\rangle, \\
&\hat{n} |n\rangle = \Bigl( J_z+ \frac{N}{2} \Bigr) |n\rangle =n |n\rangle.
\end{split}
\label{eq:operation}
\end{equation}

Here, we define the state vector corresponding to the probability distribution $P(n,t)$ as
\begin{equation}
|\phi(t) \rangle =\sum_{n=0}^{N} P(n,t) |n\rangle.
\end{equation}
As we will see, this state is nothing but the probability generating function.
Then, using the action of the operators Eq. (\ref{eq:operation}), we can rewrite the master equation to a time-evolution equation of this state vector $|\phi(t)\rangle$:
\begin{equation}
\frac{ \mathrm{d} }{ \mathrm{d} t} |\phi(t)\rangle =\hat{H}(t) |\phi(t)\rangle,
\label{eq:time evolution}
\end{equation}
\begin{equation}
\hat{H}(t)= J_+ f(\hat{n}) - (N-\hat{n})f(\hat{n}) + J_- g(\hat{n}) -\hat{n} g(\hat{n}),
\label{eq:Hamiltonian}
\end{equation}
where $\hat{I}$ is the identity operator and we have defined the Hamiltonian $\hat{H}$ of the time evolution.

The formal solution of the time evolution equation Eq. (\ref{eq:time evolution}) is given by
\begin{align}
|\phi(t)\rangle =& T( e^{\int_0^t \hat{H}(t') \mathrm{d}t'} ) |\phi(0)\rangle \notag \\
\equiv & U(t) |\phi(0) \rangle,
\end{align}
where $T$ denotes the time-ordered product and $U(t)$ is a time-evolution operator satisfying the following differential equation
\begin{equation}
\frac{ \mathrm{d} }{ \mathrm{d} t} U(t)  =\hat{H}(t) U(t). \\
\end{equation}

\subsection{Probability generating function}
In this section, we discuss how various statistics of the probability distribution can be calculated from the state vector $|\phi(t) \rangle$.
In quantum mechanics, the expectation value of a physical quantity is given by an inner-product of the corresponding operator with respect to the state vector.
On the other hand in stochastic system, the statistics is given by an inner-product of the corresponding operator with respect to the projection vector, as defined in the following.
First, noting the orthonormalization relation $\langle 0| \frac{1}{m!} (J_- )^m |n\rangle =\delta_{nm}$, the normalization condition of the probability distribution is calculated by the following inner-product,
\begin{align}
\sum_{n=0}^N P(n,t) =&\sum_{n=0}^N \sum_{m=0}^N P(n,t) \delta_{nm} \notag \\
=& \sum_{n=0}^N \sum_{m=0}^N \langle 0| \frac{1}{m!} (J_-)^m P(n,t) |n\rangle \notag \\
=& \langle \mathcal{P} |\phi(t)\rangle,
\end{align}
where
\begin{equation}
|\mathcal{P}\rangle =e^{J_+}|0\rangle
\end{equation}
is the projection vector.
Then given the state vector $|\phi(t)\rangle$, the expectation value of certain function $q(n)$ is given by the inner-product of the corresponding operator $q(\hat{n})$ as
\begin{equation}
\sum_{n=0}^N  q(n) P(n,t)= \langle \mathcal{P}|q(\hat{n}) |\phi(t)\rangle.
\end{equation}
For example, the probability generating function $G(x)$, which corresponds to $q(n) = x^n$, is calculated as 
\begin{align}
G(x)=&\langle \mathcal{P}|x^{\hat{n}} |\phi(t)\rangle \notag \\
=& \sum_{n=0}^N \sum_{m=0}^N \langle 0| \frac{1}{m!} (x J_-)^m P(n,t) |n\rangle \notag \\
=&\langle 0|e^{x J_-} |\phi(t)\rangle,
\end{align}
where the operator $x^{\hat{n}}$ is defined by the Taylor expansion, $x^{\hat{n}} =\sum_{m=0}^{\infty} (\ln x)^m \hat{n}^m$.
In the same way, the moment generating function, which corresponds to $q(n) = e^{k n}$, is calculated as
\begin{align}
M(k)=&\langle \mathcal{P}|e^{k \hat{n}} |\phi(t)\rangle \notag \\
=&\langle 0|e^{e^k J_-} |\phi(t)\rangle.
\end{align}

Seeing from a different perspective, the above calculation indicates that the state vector $|\phi(t)\rangle$ is nothing but the probability generating function.
In fact, if we define the unnormalized spin coherent state as
\begin{equation}
|\;\!\!| x\rangle \equiv e^{x J_+} |0\rangle,
\end{equation}
then, the coherent state representation of the state vector $|\phi(t)\rangle$ is given by
\begin{equation}
\langle x|\;\!\!| \phi(t) \rangle = \sum_{n=0}^N P(n,t) x^n.
\end{equation}
This is the definition of the probability generating function.

Here, we mention the difference between our su(2) method and that in the previous study. 
In the previous study, a non-Hermitian representation of the su(2) operators was used to realize the operator action Eq. (\ref{eq:operation})~\cite{Tailleur}, instead of the modification of the state normalization.
That is, the lowering operator is not Hermite conjugate to the raising operator, $(J_+ )^{\dagger} \ne J_-$.
This non-Hermiticity makes a complicated situation that the coherent state $\langle z |\;\!\!| = \langle 0 | e^{\bar{z} J_-}$ is not adjoint to the coherent state $|\;\!\!| z \rangle = e^{z J_+} | 0 \rangle$.
On the other hand in our formalism, the Hermiticity is retained.
Therefore, it is anticipated that in our formalism one can readily apply the methods developed in quantum mechanics and calculation would become concise.
This is demonstrated in the following sections.

\subsection{Probability distribution and coherent state}
Let us consider the case of the binomial distribution:
\begin{align}
P(n) =& \mathrm{Bin}(n|N,\theta) \notag \\
=& {}_N C_n \theta^n (1-\theta)^{N-n}.
\end{align}
The binomial distribution is represented by the state vector $| \phi \rangle$ as
\begin{align}
|\phi \rangle =& \sum_{n=0}^N \mathrm{Bin}(n|N,\theta) \, |n\rangle \notag \\
=& \sum_{n=0}^N \frac{N!}{(N-n)! n!} \theta^n (1-\theta)^{N-n}  \frac{(N-n)!}{N!} (J_+)^n |0 \rangle \notag \\
=& (1-\theta)^N \sum_{n=0}^N \frac{1}{n!} \Bigl( \frac{\theta}{1-\theta} J_+ \Bigr)^n |0 \rangle \notag \\
=& (1+ y )^{-N} e^{y J_+} |0 \rangle,
\end{align}
where in the last line we have set $y = \frac{\theta}{1-\theta}$.
That is, the binomial distribution can be represented by an unnormalized spin coherent state.
Then when the initial state at time $t=0$ is given by the binomial distribution, the probability generating function $G(x, y, T)$ of the system at an arbitrary time $t=T$ is given as an equivalent form to the spin coherent state propagator as follows,
\begin{equation}
G(x, y,T) = (1+ y)^{-N}\langle x|\;\!\!| U(T) |\;\!\!| y \rangle .
\end{equation}

The case where the initial state has a specific number can be treated in a similar way.
In fact, since the specific number state can be represented as $ |n\rangle =\frac{(N-n)!}{N!} (J_+)^n |0\rangle$, one can express the specific number state as a derivative of the binomial state,
\begin{equation}
|n \rangle = \frac{(N-n)!}{N!} \frac{ \partial^n}{\partial y^n} (1+ y)^{N} |\;\!\!| y \rangle \Bigr|_{y=0}.
\end{equation}
Therefore, the generating function $G(x, n,T)$ with the initial state $|n\rangle$ at time $t=T$ can be expressed as
\begin{equation}
G(x, n,T) = \frac{(N-n)!}{N!} \frac{ \partial^n}{\partial y^n}   \langle x|\;\!\!| U(T) |\;\!\!| y \rangle \Bigr|_{y=0}.
\end{equation}

\subsection{Path integral representation}
Since the probability generating function is expressed as a propagator of the spin coherent state, this generating function can be expressed as a path integral form dividing the time-evolution operator (see Appendix \ref{sec:path integral})
\begin{equation}
G(x, y, T) = (1+y)^{-N} \int_{z(0) = y}^{\bar{z}(T) = x} \mathcal{D} \bar{z} \mathcal{D} z 
\exp\Bigl\{ S[\bar{z}(t), z(t)] \Bigr\}.
\end{equation}
$S[\bar{z}(t), z(t)]$ is the action functional
\begin{equation}
S[\bar{z}(t), z(t)] =  j \ln (1+x z(T) ) +j \ln (1+\bar{z}(0) y) + \int_0^T \mathrm{d} t \biggl\{ j \frac{ \dot{\bar{z}} z - \bar{z} \dot{z} }{1+\bar{z} z} + H(\bar{z}, z) \biggr\},
\end{equation}
where the overdot denotes time derivative and the variables $z(t)$ and $\bar{z}(t)$ have the following boundary conditions,
\begin{gather}
z(0) = y, \\
\bar{z}(T) = x.
\end{gather}
Hamiltonian $H(\bar{z}, z)$ is given by a matrix element of the operator $\hat{H}$ with respect to the unnormalized spin coherent state
\begin{equation}
H(\bar{z},z) \equiv \frac{\langle z |\;\!\!| \hat{H} |\;\!\!|z\rangle}{\langle z |\;\!\!| z\rangle},
\end{equation}
where $\langle z |\;\!\!| z\rangle= (1+\bar{z}z)^{2j}$.
Specific matrix elements of lower powers of operators are calculated in Appendix \ref{sec:path integral}.

\section{Application to linear SIS model \label{sec:application} }
In this section, we apply our formalism to the simplest stochastic system with finite-state-level.
This is the linear SIS epidemic model.
In this model, each individual can take either susceptible state $S$ or infected state $I$.
We take the state of the system $n(t)$ as the number of infected individuals.  
Therefore, the number of susceptible individuals is $N-n(t)$.
Transition probabilities of this model is given as follows,
\begin{equation}
\begin{cases}
f(n) = \mu(t), \\
g(n) =  \alpha(t).
\end{cases}
\end{equation}
That is, each susceptible individual is infected with a probability $\mu(t)$ per unit time, and transitions to the infected state.
On the other hand, an individual in the infected state recovers from the infection disease with a probability $\alpha(t)$ per unit time, and transitions to the susceptible state.
Immunity is not taken into account upon recover from the infection.
Thus, recovered individuals return immediately to the susceptible state.
We allow the transition probabilities $\mu(t)$ and $\alpha(t)$ to have arbitrary time-dependences for generality.
A natural specific assignment of the recover probability may be to set $\alpha$ as time-independent.
The infection probability $\mu(t)$ may have periodicity to reflect the seasonal epidemic.
Then, the corresponding master equation is given by
\begin{align}
\frac{\partial}{\partial t} P(n,t)=&(N-n+1) \mu(t) P(n-1,t)+(n+1) \alpha(t) P(n+1,t) \notag \\
&-(N-n) \mu(t) P(n,t)-n \alpha(t) P(n,t).
\end{align}
It should be noted that this is a linear model, where the allowed infection process is $S \rightarrow I$, and the process $I+S \rightarrow 2 I$ have not been considered.
Thus, our model is different to the usual SIS epidemic model~\cite{Weiss,Keeling}.
The Hamiltonian corresponding to the master equation is given as a linear combination of the su(2) operators,
\begin{equation}
\hat{H}(t)= -(N \hat{I}-\hat{n}) \mu(t) +\mu(t) J_+ - \alpha \hat{n} +\alpha J_-.
\label{eq:SIS Hamiltonian}
\end{equation}

\subsection{Wei-Norman method}
Here, we analyze the linear SIS model by an algebraic method.
A similar attempt has been already made in the previous study, however the analysis has been restricted to a specific initial state or a specific transition probability~\cite{Shang}.
Therefore, we make a complete analysis here.

From the algebraic property of this system, the time-evolution operator $U(t)$ can be expressed as a decomposed exponential form (see Appendix \ref{Wei-Norman})~\cite{Wei,Wei2}:
\begin{equation}
U(t)=e^{g_0(t) \hat{I} } e^{g_1(t) J_- } e^{g_2(t) J_z}  e^{g_3(t) J_+},
\end{equation}
where each function $g_i (t)$ is respectively given by
\begin{equation}
g_0 (t) = -\frac{N}{2} \int_0^t \bigl( \alpha(t) +\mu(t) \bigr) \mathrm{d}t',
\end{equation}
\begin{equation}
\begin{split}
g_1 (t) =&-1+ \bigl(1+ W(t) \bigr)^{-1} e^{-\mathcal{D} (t)},
\end{split}
\end{equation}
\begin{equation}
\begin{split}
g_2 (t) =& D(t) +2 \ln \bigl(1+ W(t) \bigr),
\end{split}
\end{equation}
\begin{equation}
\begin{split}
g_3 (t) =&1 - \bigl( 1+ W(t) \bigr)^{-1}, 
\end{split}
\end{equation}
\begin{gather}
W(t) \equiv  \int_0^t \mu(t) e^{-D(t')}  \mathrm{d}t', \\
D(t)  \equiv -\int_0^t \bigl( \alpha(t) + \mu(t) \bigr) \mathrm{d}t'.
\end{gather}

\subsubsection{Binomial initial state}
We first consider the case of the binomial initial state at initial time $t=0$,
\begin{align}
|\phi(0) \rangle =& \sum_{n=0}^N \mathrm{Bin}(n|N,\theta_i) \, |n\rangle \notag \\
=& (1+ y )^{-N} e^{y J_+} |0 \rangle,
\end{align}
where $y = \frac{\theta_i}{1-\theta_i}$.
By using formula of su(2) operators (see Appendix \ref{sec:operator}), the probability generating function $G(x, y, T)$ at time $t=T$ is calculated as follows,
\begin{align}
G(x, y, T)=& (1+y)^{-N} \langle x |\;\!\!|  U(T) |\;\!\!| y \rangle  \notag \\
=& (1+y)^{-N}  \langle 0| e^{x J_-}   e^{g_0 \hat{I}} e^{ g_1 J_-} e^{ g_2 J_z} e^{g_3 J_+} e^{y J_+} |0 \rangle \notag \\
=& (1+y)^{-N} e^{g_0}  \langle 0| e^{(x +g_1) J_-}  e^{g_2 J_z} e^{(y +g_3 ) J_+} |0 \rangle \notag \\
=& (1+y)^{-N} e^{g_0}  \langle 0| e^{(x +g_1) J_-}  e^{(y+g_3) e^{g_2} J_+ } e^{ g_2 J_z} |0 \rangle  \notag \\
=& (1+y)^{-N} e^{g_0}e^{ -J g_2}   \langle 0|  e^{(y + g_3) e^{g_2} [J_+ -2(x +g_1) J_z -(x +g_1)^2 J_-]  }  e^{(x +g_1) J_-} |0 \rangle \notag \\
=& (1+y)^{-N} e^{g_0}e^{ -J g_2}   \bigl[1+ (x+g_1) (y + g_3) e^{g_2} \bigr]^{N},
\end{align}
where in the third and fourth line we have used the ordering formulae Eqs. (\ref{eq:order z+}), (\ref{eq:order -+}), and used the decomposition formula Eq. (\ref{eq:decomposition}) in the last line. 
Using the explicit expressions for $g_i(t)$, we finally arrive at the following expression for the probability generating function,
\begin{align}
G(x, y)=& \Bigl[e^{ D } (1+ W -(1+y)^{-1}) x +1- e^{ D } (1+ W -(1+y)^{-1}) \Bigr]^N \notag \\
\equiv& \Bigl[e^{ D } ( W +\theta_i) x +1- e^{ D } ( W +\theta_i) \Bigr]^N.
\end{align}
This is the probability generating function of the binomial distribution with the rate parameter given by $\theta(t)=e^{ D } ( W +\theta_i)$.
Thus, when the initial distribution is given by the binomial distribution, probability distribution retains the shape of the binomial distribution, and time-dependence in the transition probabilities only affects its rate parameter to be time-dependent. 
This result seems plausible, reflecting the fact that the Hamiltonian Eq.~(\ref{eq:SIS Hamiltonian}) takes the most general form of the coherence-preserving Hamiltonian in quantum mechanics, where the minimum uncertainty of the quantum fluctuation is retained in the course of a time-evolution.
The moment generating function can also be calculated in the same way, changing $x$ in the probability generating function to $e^k$,
\begin{equation}
M(k, y)= \Bigl[e^{ D } ( W +\theta_i) e^k  +1- e^{ D } ( W +\theta_i) \Bigr]^N.
\end{equation}

\subsubsection{Specific number initial state}
Next, we consider the case where the initial state is a specific number state $| m\rangle$.
In this case, the generating function is obtained by differentiating that of the binomial initial state
\begin{align}
G(x, m, T)
=& \frac{(N-m)!}{N!} \frac{\partial^m}{\partial y^m}
 \Bigl[e^{ D } ((1+ W ) y+ W ) x +1- e^{ D } ((1+ W) y+ W ) \Bigr]^N \Bigr|_{y=0} \notag \\
=& \Bigl[e^{ D } (1+ W ) x +1- e^{ D } (1+ W ) \Bigr]^m
\Bigl[e^{ D } W  x +1- e^{ D } W  \Bigr]^{N-m} \notag \\
\equiv & \bigl[ \theta_1 x +1- \theta_1 \bigr]^m \bigl[ \theta_0 x +1- \theta_0 \bigr]^{N-m},
\end{align}
where we have defined two rate parameters of the binomial states, $\theta_0 = e^{ D } W$ and $\theta_1 = e^{ D } (1+ W )$.
The generating function consists of the product of that of the binomial distributions.
If a sufficient time passes and information on the initial state vanishes, the probability generating function converges to that of a single binomial distribution.
\begin{equation}
G(x, m )= \bigl[ \theta_0 x +1- \theta_0 \bigr]^{N},
\end{equation}

Now we calculate the probability distribution explicitly.
This is done by dividing cases according to the magnitude relation of $\{m, \;  N-m , \; n\}$.
The results are the following:
\begin{equation}
\begin{split}
& P(n,t) \\
&= \begin{dcases}
& \!\!\!\!\!\!\!\! \sum_{l=0}^n { m \choose l } \theta_1^l (1-\theta_1)^{m-l} { N-m \choose n-l }  \theta_0^{n-l} (1-\theta_0)^{N-m-n+l}, \;\;\;
                    [ n \le m, \;n \le N-m ],  \\
& \!\!\!\!\!\!\!\! \sum_{l=0}^{N-m} { m \choose n-l }   \theta_1^{n-l}  (1-\theta_1)^{m-n+l}{ N-m \choose l }  \theta_0^{l}  ( 1-\theta_0)^{N-m-l}, \:
                    [ N-m < n \le m ],  \\
& \!\!\!\!\!\!\!\! \sum_{l=0}^{m} { m \choose l } \theta_1^{l} (1-\theta_1)^{m-l} { N-m \choose n-l }  \theta_0^{n-l} (1-\theta_0)^{N-m-n+l}, \;\;\;\;\;\;\;\;
                   [ m < n ].
\end{dcases}
\end{split}
\end{equation}

\subsection{Classical approximation to path integral}
In this section, we derive the probability generating function by evaluating the path integral directly to check the consistency of our formalism.
Here, we consider the case with time-independent transition probability for simplicity.
The generating function has a following path integral form,
\begin{equation}
G(x, y, T) = (1+y)^{-N} \int_{z(0) = y}^{\bar{z}(T) = x} \mathcal{D} \bar{z} \mathcal{D} z 
\exp\Bigl\{ S[\bar{z}(t), z(t)] \Bigr\}.
\end{equation}
The matrix element of the Hamiltonian $H(\bar{z}, z)$ in the action corresponding to Eq. (\ref{eq:SIS Hamiltonian}) is given by (see Appendix \ref{sec:path integral}),
\begin{equation}
H(\bar{z},z) = 2j \frac{(1-\bar{z})}{(1+\bar{z}z) } (-\mu+\alpha z).
\end{equation}
Now we consider the ``classical'' trajectories of the variables $z(t)$, $\bar{z}(t)$ of this system, which correspond to the extremal trajectories of the action.
These variables obey the following Hamilton's equations respectively,
\begin{gather}
\dot{z} =  \mu +(\mu-\alpha)z -\alpha z^2, \\
\dot{\bar{z}} = -\alpha- (\mu-\alpha) \bar{z}+\mu \bar{z}^2.
\end{gather}
This the Riccati differential equation, therefore the solutions satisfying the boundary conditions $z(0) = y$ and $\bar{z}(T) = x$ are respectively given by,
\begin{align}
z(t) = &-1+\frac{e^{wt}}{\frac{\alpha}{w}(e^{wt}-1)+\frac{1}{y+1}} \notag \\
=& -1+\frac{1}{\alpha}\frac{\mathrm{d} }{\mathrm{d}t} \ln \Bigl[\frac{\alpha}{w} (e^{wt}-1)+\frac{1}{y+1}\Bigr]
\end{align}
\begin{align}
\bar{z}(t) =&1+\frac{e^{wt}}{\frac{\mu}{w}(e^{wT}-e^{wt})+\frac{1}{x-1} e^{wT}} \notag \\
=& 1-\frac{1}{\mu }\frac{\mathrm{d} }{\mathrm{d} t} \ln \Bigl[\frac{\mu}{w}(e^{wT}-e^{wt})+\frac{1}{x-1} e^{wT}\Bigr]
\end{align}

The ``classical'' action is obtained by inserting the ``classical'' solutions into the action functional,
\begin{align}
S_{\mathrm{cl}}(x, y, T)
=&j \ln(1+x z(T)) + j \ln(1+\bar{z}(0) y)  + j \int_0^T \mathrm{d}t ( -2\mu +\alpha z + \mu \bar{z} ) \notag \\
=&  2j \ln \biggl\{ (x-1) \Bigl[ \frac{\mu}{w} (e^{wT}-1)+\frac{y}{1+y} \Bigr] e^{-wT} +1 \biggr\} + 2j\ln(1+y)
\end{align}
Using the ``classical'' action, we obtain the ``classical'' approximation to the probability generating function,
\begin{align}
G_{\mathrm{cl}}(x, y, T) \equiv & (1+y)^{-N} \exp \Bigl\{ S_{\mathrm{cl}}(x,y,T) \Bigr\} \notag \\
=& \biggl\{ (x-1) \Bigl[ \frac{\mu}{w} (e^{wT}-1)+\frac{y}{1+y} \Bigr] e^{-wT} +1 \biggr\}^{N}.
\end{align}
This is precisely the generating function of the binomial distribution.
Thus, we have confirmed that the ``classical'' approximation gives the exact result in the linear SIS model, $G(x,y,T) = G_{\mathrm{cl}}(x,y,T)$, and that our path integral representation gives the correct result.

\section{Conclusion \label{sec:conclusion} }
In this paper, we propose the method to analyze a discrete stochastic process with finite-state-level.
By using a representation of su(2) algebra, the master equation of the system is rewritten to a time evolution equation for the state vector corresponding to the probability generating function.
Then, we found that the generating function can be expressed as a propagator of the spin coherent state.
Furthermore, the generating function has a path integral representation of the spin coherent state.
Also, we found the correspondence between coherent state and the probability distribution.
That is, the generating function of the binomial distribution is expressed by the spin coherent state.
As a simplest application of our method, a linear SIS epidemic model with arbitrary time-dependent transition probabilities was analyzed.
The generating function of the system was calculated using underling algebraic property of the system or a path integral representation.
We considered two initial states, the binomial state and the specific number state.
After a sufficient time passed, the generating function converges to that of the binomial distribution in both cases, where the time-dependent transition probabilities only affect the rate parameter of the binomial distribution.

The model analyzed in this paper is the too simple system in which the interactions between each individual are ignored, and the result is little bit trivial.
However, our results indicate that our formalism is suited for analyzing the finite-level system and gives basis for applying the method developed in quantum mechanics to more complicated stochastic systems.
One example of such an approximation scheme is the Wentzel-Kramers-Brillouin (WKB) approximation to the propagator using a spin coherent state path integral~\cite{WKB,Sacramento}.
It is of significance to see how such an approximation scheme captures the finite-size effect.
In addition, it is of interest to further investigate the relationship between probability distributions and generalized coherent states.
The relationship will be helpful in analyzing the corresponding stochastic systems as illustrated in this paper.

\appendix
\section{Path integral representation \label{sec:path integral}}
In this appendix, we derive the spin coherent state path integral representation to the probability generating function in detail for the purpose of specifying our notation.
In the path integral formalism, one first divides the time evolution operator $U(t)$ into that of the short-time interval,
\begin{equation}
U(t) = \lim_{N \rightarrow \infty} \Bigl( 1 + \frac{ \hat{H} T}{N} \Bigr)^N.
\end{equation}
Then, inserting the $(N-1)$ over-completeness relations in terms of spin coherent state between the $N$ short-term time evolution operators, 
\begin{equation}
\hat{ I } = \frac{2j+1}{\pi} \int_{-\infty}^{\infty} \frac{ \mathrm{d} \bar{z} \mathrm{d}z}{(1+\bar{z}z)^{2j+2} } |\;\!\!|z\rangle \langle z|\;\!\!|,
\end{equation}
the probability generating function can be expressed as product of $N$ short-term propagators as follows,
\begin{align}
G(x, y, T) =& (1+y)^{-N} \lim_{N \rightarrow \infty} \biggl[ \prod_{i=1}^{N-1} \frac{2j+1}{\pi} \int \frac{\mathrm{d} \bar{z}_i \mathrm{d} z_i}{(1+\bar{z}_i z_i)^{2j+2}} \biggr] \notag \\
&
\langle z_N |\;\!\!|\Bigl(1+\frac{ \hat{H} T}{N} \Bigr) |\;\!\!| z_{N-1} \rangle \cdots 
\langle z_k |\;\!\!| \Bigl(1+\frac{ \hat{H} T}{N} \Bigr) |\;\!\!| z_{k-1} \rangle \cdots
\langle z_1 |\;\!\!| \Bigl(1+\frac{ \hat{H} T}{N} \Bigr) |\;\!\!| z_0\rangle \notag \\
=& (1+y)^{-N} \lim_{N \rightarrow \infty} \biggl[ \prod_{i=1}^{N-1} \frac{2j+1}{\pi} \int \frac{\mathrm{d} \bar{z}_i \mathrm{d} z_i}{(1+\bar{z}_i z_i)^{2}} \biggr] 
(1+\bar{z}_N z_N)^j (1+\bar{z}_0 z_0)^j \notag \\
&
\prod_{k=1}^N \frac{(1+\bar{z}_k z_{k-1})^{2j} }{ (1+\bar{z}_k z_k)^j (1+\bar{z}_{k-1} z_{k-1} )^j } \Bigl(1+\frac{H(\bar{z}_k, z_{k-1}) T}{N} \Bigr),
\end{align}
where
\begin{equation}
H(\bar{z}_k, z_{k-1}) \equiv \frac{\langle z_k |\;\!\!| \hat{H} |\;\!\!| z_{k-1} \rangle }{\langle z_k |\;\!\!| z_{k-1} \rangle },
\end{equation}
is a matrix element of the Hamiltonian, and we have set $ z_N \equiv x$, $z_0 \equiv y$.
Here, if we assume $\Delta  z_k = z_k - z_{k-1}$ as a small quantity of order $\mathcal{O}(T/N)$, the following approximation can be made,
\begin{equation}
\frac{(1+\bar{z}_k z_{k-1})^{2j} }{ (1+\bar{z}_k z_k)^j (1+\bar{z}_{k-1} z_{k-1} )^j } \simeq 1+ j\frac{\Delta \bar{z}_k z_{k-1} - \Delta z_k \bar{z}_{k-1}}
{1+\bar{z}_{k-1} z_{k-1} }.
\end{equation}
Re-expressing the above $\mathcal{O} (T/N)$ terms into the exponential function, we finally arrive at the following path integral representation to the probability generating function,
\begin{equation}
G(x, y, T) = (1+y)^{-N} \int_{z(0) = y}^{\bar{z}(T) = x} \mathcal{D} \bar{z} \mathcal{D} z 
\exp\Bigl\{  S[\bar{z}(t), z(t)] \Bigr\},
\end{equation}
where the integration measure is defined by
\begin{equation}
\int \mathcal{D} \bar{z} \mathcal{D} z \equiv \lim_{N \rightarrow \infty} \biggl[ \prod_{i=1}^{N-1} \frac{2j+1}{\pi} \int \frac{\mathrm{d} \bar{z}_i \mathrm{d} z_i}{(1+\bar{z}_i z_i)^{2j+2}} \biggr].
\end{equation}
The action functional is given by
\begin{equation}
S[\bar{z}(t), z(t)] = j \ln (1+x z(T) ) +j \ln (1+\bar{z}(0) y) + \int_0^T \mathrm{d} t \biggl\{ j \frac{ \dot{\bar{z}} z - \bar{z} \dot{z} }{1+\bar{z} z} + H(\bar{z}, z) \biggr\},
\end{equation}
where the variables $z(t)$ and $\bar{z}(t)$ have the boundary conditions $z(0) = y$ and $\bar{z}(T) = x$.
The Hamiltonian $H(\bar{z}, z)$ is defined as a matrix element of the operator $\hat{H}$.
For example, matrix elements lower powers of operators are calculated as follows.
The first-order operators are respectively given by
\begin{gather}
\frac{\langle z_1|\;\!\!| J_z |\;\!\!| z_2\rangle}{\langle z_1 |\;\!\!| z_2\rangle} =\frac{ -j (1-\bar{z}_1 z_2) }{ (1+\bar{z}_1 z_2) }, \\
\frac{\langle z_1|\;\!\!| J_+ |\;\!\!| z_2\rangle}{\langle z_1 |\;\!\!| z_2\rangle} =\frac{ 2j \bar{z}_1 }{ (1+\bar{z}_1 z_2) }, \\
\frac{\langle z_1|\;\!\!| J_- |\;\!\!| z_2\rangle}{\langle z_1 |\;\!\!| z_2\rangle} =\frac{ 2j z_2 }{ (1+\bar{z}_1 z_2) }.
\end{gather}
The second-order operators are respectively given by
\begin{gather}
\frac{\langle z_1|\;\!\!| J_+^2 |\;\!\!| z_2\rangle}{\langle z_1 |\;\!\!| z_2\rangle} =\frac{ 2j (2j-1) \bar{z}_1^2 }{ (1+\bar{z}_1 z_2)^2 }, \\
\frac{\langle z_1|\;\!\!| J_-^2 |\;\!\!| z_2\rangle}{\langle z_1 |\;\!\!| z_2\rangle} =\frac{ 2j (2j-1) z_2^2 }{ (1+\bar{z}_1 z_2)^2 }, \\
\frac{\langle z_1|\;\!\!| J_+  J_-  |\;\!\!| z_2\rangle}{\langle z_1 |\;\!\!| z_2\rangle} =\frac{ 4j^2 \bar{z}_1 z_2 +2j \bar{z}_1^2 z_2^2  }{ (1+\bar{z}_1 z_2)^2 }, \\
\frac{\langle z_1|\;\!\!| J_- J_+  |\;\!\!| z_2\rangle}{\langle z_1 |\;\!\!| z_2\rangle} =  \frac{ 4j^2 \bar{z}_1 z_2 +2j  }{ (1+\bar{z}_1 z_2)^2 }, \\
\frac{\langle z_1|\;\!\!| J_z^2 |\;\!\!| z_2\rangle}{\langle z_1 |\;\!\!| z_2\rangle} =\frac{ j^2 (1-\bar{z}_1 z_2)^2 + 2j \bar{z}_1 z_2  }{ (1+\bar{z}_1 z_2)^2 }, \end{gather}
\begin{gather}
\frac{\langle z_1|\;\!\!| J_+ J_z  |\;\!\!| z_2\rangle}{\langle z_1 |\;\!\!| z_2\rangle} = \frac{ -2j^2 \bar{z}_1 +2j (j-1) \bar{z}_1^2 z_2   }{ (1+\bar{z}_1 z_2)^2 }, \\
\frac{\langle z_1|\;\!\!| J_- J_z  |\;\!\!| z_2\rangle}{\langle z_1 |\;\!\!| z_2\rangle} = 
\frac{ 2j^2 z_2   }{ (1+\bar{z}_1 z_2) }
- \frac{   2j(2j-1) z_2   }{ (1+\bar{z}_1 z_2)^2 }.
\end{gather}

``Classical'' trajectories, which correspond the extremal trajectories of the action, obey the following Hamilton's equations,
\begin{gather}
\dot{z} = \frac{ (1+\bar{z}z)^2}{2j} \frac{\partial H }{\partial \bar{z}} , \\
\dot{\bar{z}} = -\frac{(1+\bar{z}z)^2}{2j}  \frac{\partial H}{\partial z}.
\end{gather}

\section{Wei-Norman theorem for su(2) algebra \label{Wei-Norman}}
Time evolution operator $U(t)$ obeys the following differential equation,
\begin{align}
\frac{\mathrm{d} }{\mathrm{d}  t} U(t)=&\hat{H}(t) U(t) \notag \\
                      =& \sum_{i=0}^3 a_i(t) H_i \,U(t),
\label{eq:unitary evolution}
\end{align}
where we have set the coefficients $a_i(t)$ and the operators $H_i$ as,
\begin{equation}
\begin{split}
&a_0 (t)=-\frac{N}{2} \bigl( \alpha(t)+\mu(t) \bigr), \hspace{0.5cm} a_1(t) =  \alpha (t), \\
&a_2(t)=  -\alpha (t)+\mu(t), \hspace{1.35cm} a_3 (t) =\mu (t), \\
& H_0 = \hat{I}, \hspace{0.5cm} H_1 = J_-, \hspace{0.5cm} H_2 = J_z, \hspace{0.5cm} H_3 = J_+.
\end{split}
\end{equation}
Wei-Norman theorem states that the solution of the above differential equation takes a product of exponential operators~\cite{Wei,Wei2},
\begin{equation}
U(t)=e^{g_0 (t) H_0} e^{g_1 (t) H_1} e^{g_2 (t) H_2} e^{g_3 (t) H_3},
\label{eq:Wei-Norman}
\end{equation}
where $g_i(t)$ satisfy the initial conditions $g_i(0)=0$.
The functions $g_i(t)$ can be derived by differentiating Eq. (\ref{eq:Wei-Norman}) with respect to time $t$.
In fact, the time derivative of the time-evolution operator Eq. (\ref{eq:Wei-Norman}) takes the following form,
\begin{equation}
\left( \frac{\mathrm{d} }{\mathrm{d}t} U(t) \right) U^{-1}(t)=\sum_{i=0}^3 \dot{g}_i (t) e^{g_0 (\mathrm{ad} H_0)} e^{g_1 (\mathrm{ad} H_1)} \cdots e^{g_{i-1} (\mathrm{ad} H_{i-1})} H_i,
\end{equation}
with the aid of the Baker-Campbell-Hausdorff formula,
\begin{equation}
e^{H_i} H_j e^{-H_i}= e^{(\mathrm{ad} H_i) } H_j,
\end{equation}
where the adjoint operator $\mathrm{ad} H_i$ is defined as
\begin{align}
(\mathrm{ad} H_i) H_j=&[H_i, H_j] \notag \\
=& H_i H_j -H_j H_i.
\end{align}
Using the su(2) adjoint operators,
\begin{gather}
e^{g (\mathrm{ad} J_-)} J_z = J_z +g J_-, \\
e^{g (\mathrm{ad} J_-)} J_+ =J_+ -2g J_z  -g^2 J_- , \label{eq:adjoint1} \\
e^{g (\mathrm{ad} J_z )} J_+ =e^{g} J_+, 
\label{eq:adjoint2}
\end{gather}
we obtain
\begin{equation}
\left( \frac{\mathrm{d} }{\mathrm{d} t} U(t) \right) U^{-1}(t)
=\bigl( \dot{g}_0 \bigr) \hat{I} + \bigl( \dot{g}_1 +\dot{g}_2 g_1-\dot{g}_3 e^{g_2} g_1^2 \bigr) J_-  + \bigl( \dot{g}_2 -2 \dot{g}_3 e^{g_2} g_1 \bigr)J_z + \bigl( \dot{g}_3 e^{g_2} \bigr) J_+.
\end{equation}
Comparing the above equation to the original equation Eq.~(\ref{eq:unitary evolution}), we find that $g_i$ are the solutions of the following differential equations
\begin{gather}
\dot{g}_0 = a_0, \\
\dot{g}_1 =  a_1 -a_2 g_1 -a_3 g_1^2, \\
\dot{g}_2 = a_2 +2 a_3 g_1, \\
\dot{g}_3 = a_3 e^{-g_2}. 
\end{gather}
The differential equation for $g_1(t)$ is the Riccati differential equation.
Therefore, the solutions are respectively given by,
\begin{equation}
g_0 (t) = \int_0^t a_0(t') \mathrm{d} t',
\end{equation}
\begin{equation}
g_1 (t) = -1+ \bigl( 1+ W(t) \bigr)^{-1} e^{-D(t)},
\end{equation}
\begin{equation}
g_2 (t) = D(t) +2 \ln \bigl(1+ W(t) \bigr),
\end{equation}
\begin{equation}
g_3 (t) = - \bigl( 1+ W(t) \bigr)^{-1} +1,
\end{equation}
where
\begin{equation}
W(t)= \int_0^t a_3 (t') e^{-D(t')}  \mathrm{d} t' ,
\end{equation}
\begin{equation}
D(t) =\int_0^t \bigl( a_2 (t')- 2 a_3 (t') \bigr) \mathrm{d} t' ,
\end{equation}
and we have taken the integration constants so as to satisfy the initial conditions $g_i(0)=0$.

\section{Various formulae of su(2) operator \label{sec:operator}}
\subsection{Decomposition formula}
Operators satisfying the su(2) algebra $[J_z, J_{\pm}]=\pm J_{\pm}$, $[J_+, J_-]= 2 J_z$ have a following decomposition formula~\cite{Decomposition},
\begin{equation}
\exp (u_1 J_+ + u_2 J_z + u_3 J_-) =\exp (v_1 J_+) \exp ( (\log v_2) J_z) \exp (v_3 J_-),
\label{eq:decomposition}
\end{equation}
where the parameter constants are respectively given by
\begin{equation}
v_1= \frac{(u_1/\phi) \sinh \phi}{\cosh \phi -(u_2/2\phi) \sinh \phi},
\end{equation}
\begin{equation}
v_2= \frac{1}{ [\cosh \phi -(u_2/2\phi) \sinh \phi ]^2},
\end{equation}
\begin{equation}
v_3= \frac{(u_3/\phi) \sinh \phi}{\cosh \phi -(u_2/2\phi) \sinh \phi},
\end{equation}

\begin{equation}
\phi=\sqrt{(u_2/2)^2+u_1 u_3}.
\end{equation}

\subsection{Baker-Campbell-Hausdorff lemma}
su(2) operators have the exchanging formulae.
In fact, using the formula for exponential adjoint operators Eqs. (\ref{eq:adjoint1}), (\ref{eq:adjoint2}), exponential operators can be exchanged as follows,
\begin{align}
e^{a J_z} e^{b J_+}=&e^{e^{a (\mathrm{ad} J_z) } b J_+} e^{a J_z} \notag \\
=& e^{b e^{a} J_+  } e^{a J_z},
\label{eq:order z+}
\end{align}
\begin{align}
e^{a J_-} e^{b J_+}=&e^{e^{a (\mathrm{ad} J_-) } b J_+} e^{a J_-} \notag \\
=& e^{b (J_+ -2 a J_z-a^2 J_-) } e^{a J_-}.
\label{eq:order -+}
\end{align}



\end{document}